\documentclass[iop]{emulateapj}
\usepackage{amsmath}
\usepackage{ulem}
\slugcomment{Received 2009 December 11; accepted 2010 May 19}

\begin{document}

\title{Adaptive Optics Observations of B0128+437: a Low-Mass, High-Redshift
  Gravitational Lens}

\shorttitle{AO observations of CLASS B0128+437}
\shortauthors{Lagattuta, Auger, \& Fassnacht}

\author{David J. Lagattuta\altaffilmark{1}}
  \email{lagattuta@physics.ucdavis.edu}
\author{Matthew W. Auger\altaffilmark{2}} 
\author{Christopher D. Fassnacht\altaffilmark{1}}

\altaffiltext{1}{Department of Physics, University of California,
  Davis, 1 Shields Avenue, Davis CA 95616}
\altaffiltext{2}{Department of Physics, University of California,
  Santa Barbara, CA 93106}

\begin{abstract}
We use high-resolution adaptive optics (AO) imaging on the Keck II
telescope to study the gravitational lens B0128+437 in unprecedented
detail, allowing us to resolve individual lensed quasar components
and, for the first time, detect and measure properties of the lensing
galaxy.  B0128+437 is a small separation lens with known flux-ratio
and astrometric anomalies.  We discuss possible causes for these
anomalies, including the presence of substructure in the lensing
galaxy, propagation effects due to dust and a turbulent interstellar
medium, and gravitational microlensing.  This work demonstrates that
AO will be an essential tool for studying the many new
small-separation lenses expected from future surveys.
\end{abstract}

\keywords {galaxies: high-redshift --- galaxies: individual (CLASS
  B0128+437) --- gravitational lensing: strong --- infrared: galaxies}

\section{Introduction}

Galaxies acting as strong gravitational lenses can provide important
insights into the distribution of matter on small scales, as the
brightness and position of the multiple lensed images are directly
related to the gravitational potential of the foreground galaxy.  Many
lenses are well modeled by a smooth potential.  For other lenses,
however, a simple mass model is incapable of accurately predicting the
positions of the multiple images of the background object (astrometric
anomalies), or the relative brightness between image pairs (flux-ratio
anomalies).  The existence of these astrometric and flux-ratio
anomalies in lenses has been used to study the presence of galactic
substructure \citep{dal02,xu09} -- predicted by numerical simulations
of the $\Lambda$CDM theory of structure formation \citep{die08,spr08}
-- as well as propagation effects such as dust extinction and scatter
broadening.

In this letter we discuss one such anomalous lens, CLASS B0128+437
\citep{phi00}.  First observed as part of the Cosmic Lens All-Sky
Survey \citep[e.g.,][]{mey03,bro03}, B0128+437 is a compact four-image
system with a maximum image separation of 0.54\arcsec. Very Long
Baseline Interferometry (VLBI) imaging presented by \citet{big04}
revealed three distinct subcomponents in each of the lensed source
images.  Lens modeling in B04 could not sufficiently reproduce either
the positions of these subcomponents in images B and C or the total
flux of image B, suggesting that some form of substructure was present
in the lens system.

\citet{mck04} determined a redshift of $z_s=3.124$ for the source
galaxy, and an emission line not associated with the source was
posited to be either H$\alpha$ (implying a lens redshift $z_l=0.218$),
H$\beta$ ($z_l=0.645$), or [\ion{O}{2}] ($z_l=1.145$).  Their analysis
suggested that [\ion{O}{2}] was the most likely candidate; if
confirmed, this would indicate that B0128+437 is the most distant
known gravitational lens.

Given the small image separation and a possible [\ion{O}{2}] emission
line, the lensing galaxy in the B0128+437 system may be a late-type
galaxy.  Ground based images in the infrared (IR) from UKIRT (B04)
have not been able to resolve the system, and space based images in
the IR (NICMOS; \citealt{big04b}; Figure \ref{fig:imgs}, left) and
optical (WFPC2) have had neither the resolution nor, in the case of
WFPC2, sufficient sensitivity to investigate the properties of the
system.

In this Letter we present a different method of observing B0128+437:
ground-based near-IR imaging coupled with laser guide-star adaptive
optics (LGS AO).  While AO imaging has been used on gravitational
lenses in the past \citep{cra98,mar07,mck07,aug08}, the systems
studied have been high-mass lenses producing image separations of
$\sim$2\arcsec.  With B0128+437, we present the first results of using
AO with a small-separation gravitational lens, demonstrating that AO
imaging is capable of probing the details of low-mass systems.

Throughout this Letter, we assume a concordance cosmological model,
with $\Omega_M=0.3$, $\Omega_{\Lambda}=0.7$, and a Hubble
parameter $H_0=70~h_{70}$ km s$^{-1}$ Mpc$^{-1}$. All magnitudes
presented are AB magnitudes.

\section{Observations}
We observed the B0128+437 system using the Near Infrared Camera 2
(NIRC2; K. Matthews et al., in preparation) on the Keck II telescope
along with the the LGS AO system, under photometric conditions, on UT
2009 September 12.  We used the narrow camera (10\arcsec $\times$
10\arcsec~ field of view, 0.01\arcsec/pixel scale) to properly sample
the diffraction-limited core of the point-spread function (PSF).  The
data were obtained using the Kp filter and were dithered to mitigate
the effects of bad pixels, cosmic ray strikes, and detector
persistence.  Each exposure was 180 seconds to stay within the linear
regime of the detector and to minimize the variation in sky brightness
for flat-fielding.  After visually inspecting cleaned individual
exposures and rejecting frames where the image appeared strongly
distorted due to loss of LGS lock or problems with the AO wavefront
sensors, we were left with 30 frames and a total exposure time of 5400
s.

The data were reduced using the Center for Adaptive Optics Treasure
Survey (CATS)\footnote{http://irlab.astro.ucla.edu/cats/index.shtml}
NIRC2 pipeline, modified to properly correct for the geometric
distortion of the narrow
camera\footnote{http://www.astro.caltech.edu/$\sim$pbc/AO/distortion.pdf}. The
reduced image frames were then stacked using the Drizzle routine
\citep{fru02} in IRAF, using sub-pixel offsets determined via
cross-correlation.

\begin{figure*}
\begin{center}
\centerline{
\includegraphics[width=5.8cm]{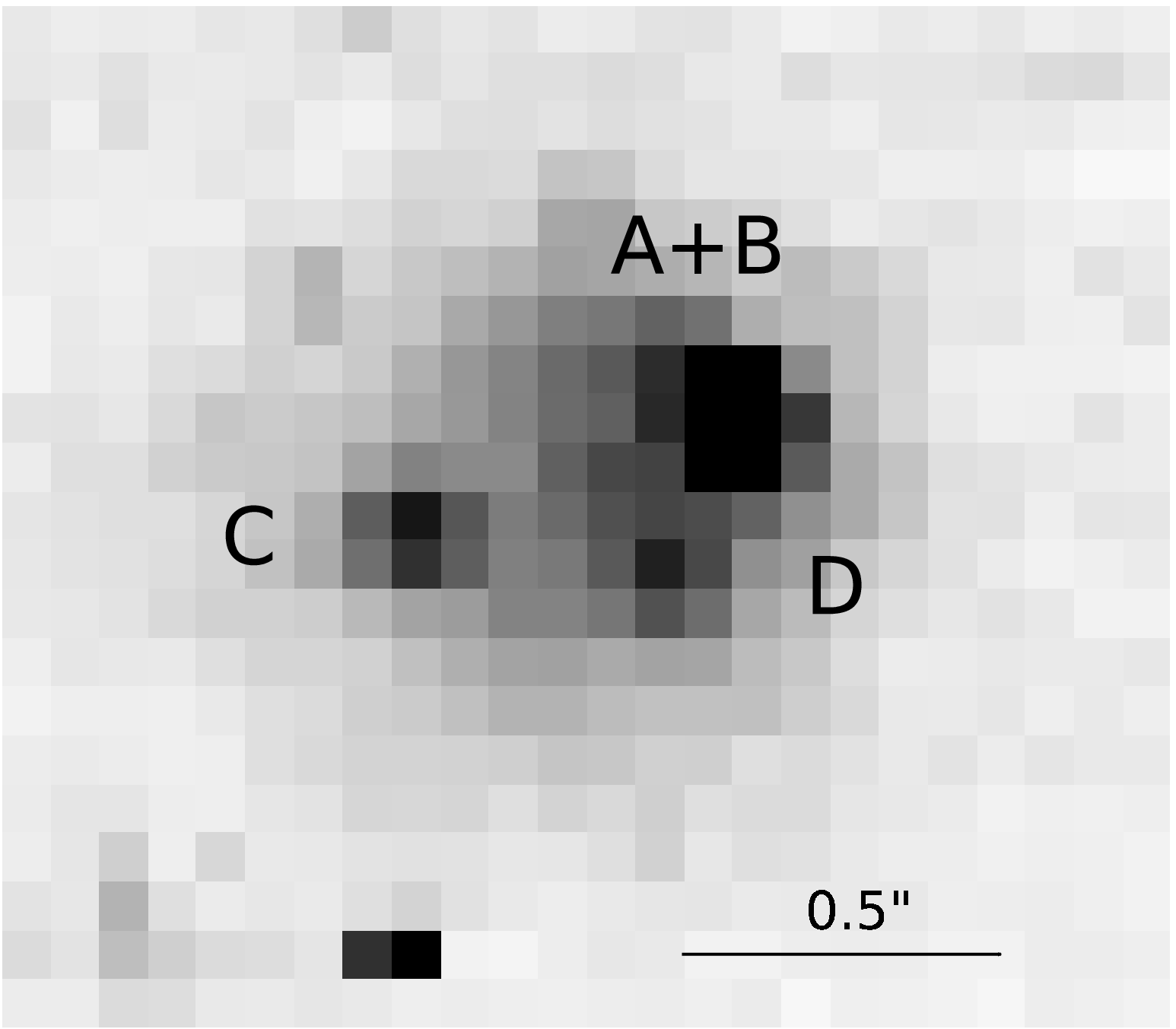}
\includegraphics[width=5.8cm]{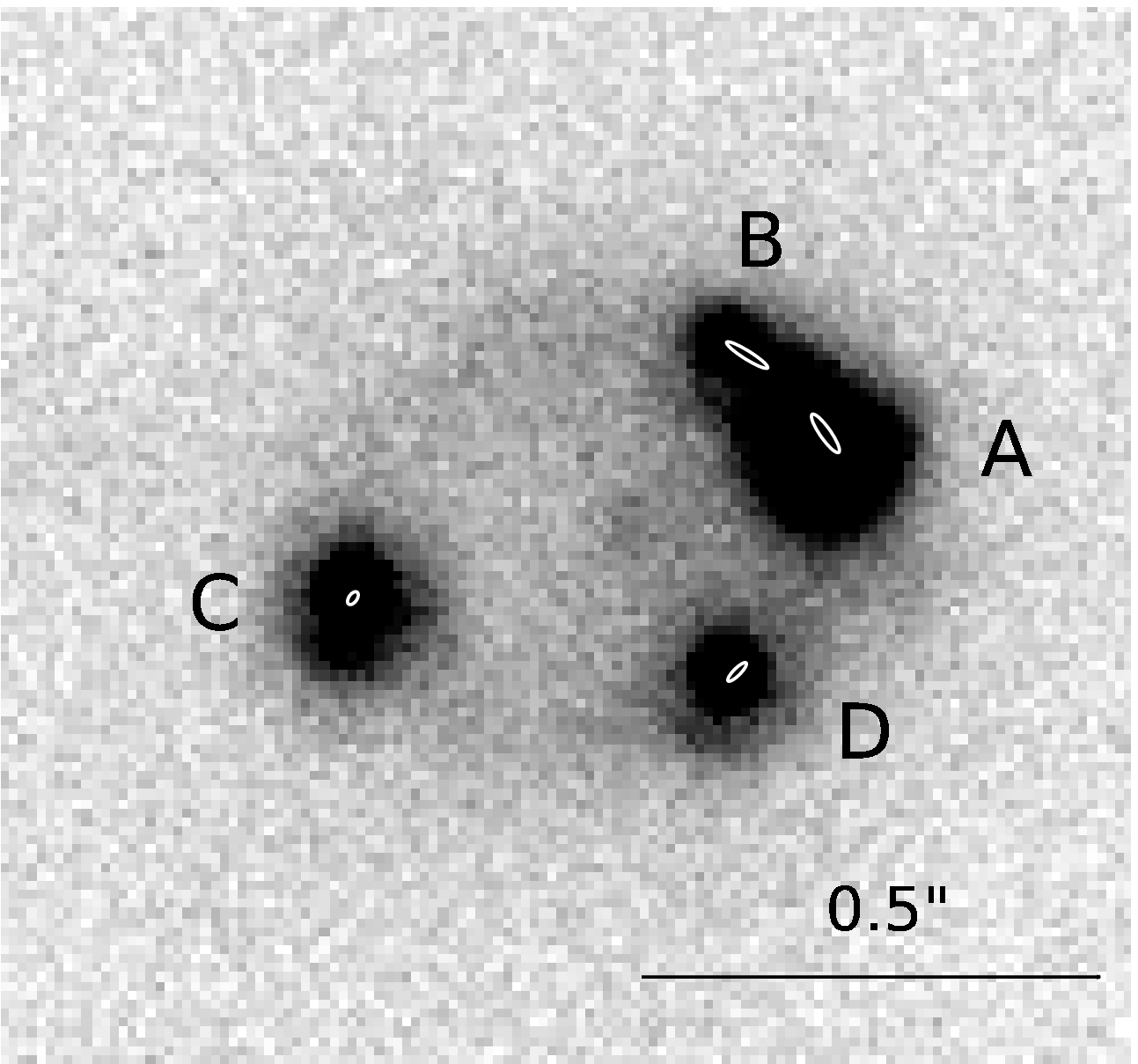}
\includegraphics[width=5.8cm]{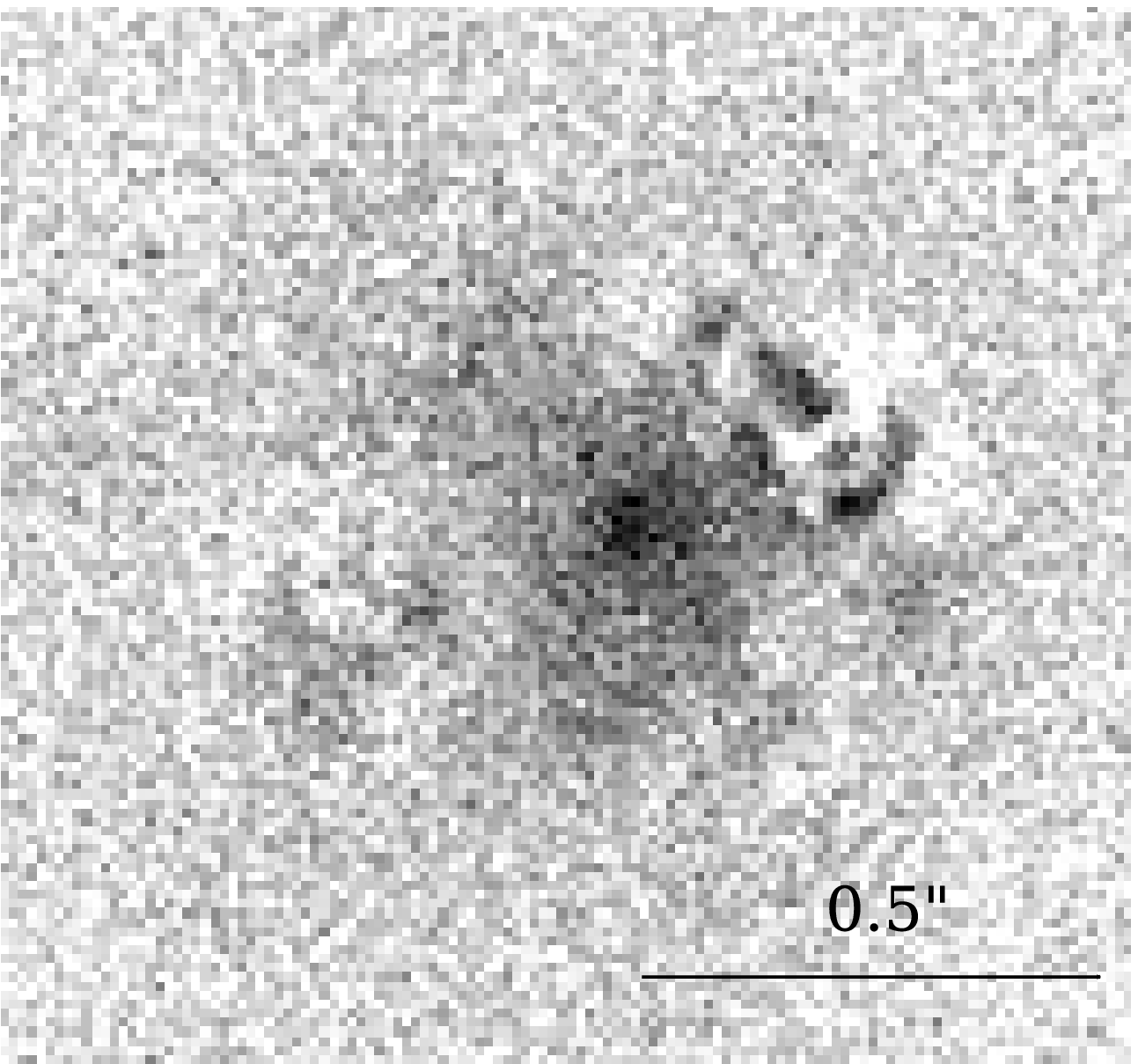}}
\end{center}
\caption{{\bf Left:} NICMOS F160W image of B0128+437.  The resolution
  of HST in this band makes it difficult to resolve individual
  components.  {\bf Middle:} AO image of B0128+437.  The four lensed
  quasar images [A{--}D] are clearly seen and easily resolved, along
  with a partial Einstein ring from the quasar host galaxy.  The
  3-$\sigma$ 2.3 GHz contours of B04 are overlayed in white.
  Additionally, the lensing galaxy can be seen lying close to A, B,
  and D in the image plane. {\bf Right:} Flux remaining in the
  B0128+437 system after subtracting the source quasar and host galaxy
  light.  The lens galaxy is now clearly visible.  Note that the image
  stretch in this panel is different from that of the middle panel.}
\label{fig:imgs}
\end{figure*}

The results of the reduction can be seen in the middle panel of Figure
\ref{fig:imgs}.  We see the four lensed quasar images clearly, along
with the lensing galaxy and a partial Einstein ring from the quasar
host galaxy.  This is a significant improvement over the 5440 s NICMOS
F160W image.

%--------------------------------------------------------------------
%--------------------------------------------------------------------

\section{Modeling Mass and Light}

%--------------------------------------------------------------------
\begin{deluxetable}{lrcc}
\tablecaption{Lens System Parameters \label{tbl:lensa}}
\tablehead{
\colhead{Component} & \colhead{Kp} &
\colhead{$\Delta\alpha$(\arcsec)\tablenotemark{a}} &
\colhead{$\Delta\delta$(\arcsec)\tablenotemark{a}}
}
\startdata
Image A   &  21.55 $\pm$ 0.03  &  $\equiv 0.0$  &  $\equiv 0.0$      \\
Image B   &  23.49 $\pm$ 0.22  &  0.099 &  0.095  \\
Image C   &  22.49 $\pm$ 0.04  &  0.521 & -0.170  \\
Image D   &  22.87 $\pm$ 0.12  &  0.109 & -0.260  \\
Lens galaxy (total)& 20.99 $\pm$ 0.08 & 0.217 & -0.104 \\ 
Lens galaxy($R_{\rm{Ein}}$)\tablenotemark{b} & 22.15 $\pm$ 0.10 & 0.217 & -0.104 \\
\enddata
\tablenotetext{a}{Uncertainties are $\sim$0.001 for B/C/D and
$\sim$0.01 for the galaxy.}
\tablenotetext{b}{Aperture magnitude with
$R=R_{\rm{Ein}}=0.24\arcsec$.}
\end{deluxetable}
%--------------------------------------------------------------------

We are primarily interested in the light of the lensing galaxy and any
substructure that might be associated with it and we therefore treat
the lensed quasar and its host galaxy (the Einstein ring) as
nuisances. We model the flux in the image as various components: a
S\'{e}rsic profile \citep{ser63} for the lensing galaxy, PSF models
for the quasi-stellar object (QSO) images, and a lensed S\'{e}rsic
profile for the QSO host galaxy.  We then marginalize over the
parameters of the source components to infer the properties of the
lensing galaxy.  Note that we do not treat the QSO images as being
lensed; this is because we suspect that simple lens models will not be
sufficient to reproduce the observed fluxes, complicating our attempts
to model the light.  We do not use the QSO image positions to
constrain the lens model, because we are able to obtain a robust
subtraction of the Einstein ring by using the data in the ring itself
and assuming the Singular Isothermal Ellipsoid (SIE) + external shear
parameters from B04 as a starting point for the lens model; the final
model is consistent with the B04 model, given the uncertainties.  It
is interesting that we are able to sufficiently remove the Einstein
ring flux with a simple model, as it implies that the known anomalies
present in the B0128+437 system are not the result of incorrect
assumptions about the global mass model, but rather due to some
localized discrepancy (\S4.2).

The PSF model is determined from the data by fitting a model of three
Gaussian components to the QSO images; one component represents the
diffraction-limited core, another represents the seeing-limited
diffuse PSF, and the third encodes structure in the PSF, as seen in
the C and D components in Figure \ref{fig:imgs} (middle).  The full
model (S\'{e}rsic lens, four QSO PSFs, and lensed QSO host) is fit to
the data iteratively, using a Levenberg-Marquardt optimization
algorithm to find the minimum in the parameter space.  A set of Markov
Chain Monte Carlo simulations are run to probe the parameter space
around this minimum and obtain uncertainty estimates for the quasar
image fluxes, lens flux, and lens position. The results of the
modeling are shown in Figure \ref{fig:imgs} (right) and the best-fit
parameters for the lensing galaxy and QSO positions and fluxes are
tabulated in Table \ref{tbl:lensa}.

\section{Discussion}
The improved image resolution provided by the AO data allows us to (1)
for the first time observationally detect the lensing galaxy and
estimate its light profile, (2) measure flux ratios between pairs of
lensed images without worrying about blending effects, and (3) search
for luminous substructure.

\subsection{The Lens Galaxy}

%--------------------------------------------------------------------
\begin{deluxetable*}{rcrrcrc}
\tablecaption{Redshift Dependent Lens Galaxy Properties \label{tbl:lensb}}
\tablehead{ 
\colhead{$z_l$} &\colhead{Stellar Age\tablenotemark{a}} & 
\colhead{$\log_{10}L_{\rm K, lens}$} &
\colhead{$\log_{10}L_{\rm K,{\rm Ein}}$} & \colhead{$\log_{10}{\rm M_{Ein}}$\tablenotemark{b}} & \colhead{M/L$_{K,\rm{Ein}}$} &
\colhead{$\sigma_{\rm SIE}$\tablenotemark{b}}\\ 
\colhead{} & \colhead{($h_{70}^{-1}~$ Gyr)} & 
\colhead{($h_{70}^{-2}~L_{\sun}$)} & \colhead{($h_{70}^{-2}~L_{\sun}$)} &   
\colhead{($h_{70}^{-1}~M_{\sun}$)} & \colhead{($h_{70}~M_{\sun}$/$L_{\sun}$)} & 
\colhead{(km s$^{-1}$)} 
}
\startdata
0.218  &10.8  & 9.53 $\pm$ 0.03 & 9.07  $\pm$ 0.04  & 9.78  & 5.13 $\pm$ 0.04 & 98\\
0.645  & 8.5  &10.55 $\pm$ 0.04 &10.09  $\pm$ 0.04  &10.20  & 1.29 $\pm$ 0.04 &114\\
1.145  & 3.2  &11.07 $\pm$ 0.05 &10.61  $\pm$ 0.05  &10.44  & 0.68 $\pm$ 0.05 &138\\

\enddata 
\tablenotetext{a}{Measured assuming a stellar fraction $f_*=1.0$ for
$z_l=\{0.645,1.145\}$ and $f_*=0.3$ for $z_l=0.218$.}
\tablenotetext{b}{Determined using the B04 mass model with $R_{\rm
Ein}=0.24\arcsec$.}
\end{deluxetable*}
%--------------------------------------------------------------------

In the rightmost panel of Figure \ref{fig:imgs} we clearly see the
B0128+437 lens galaxy.  We measure a total magnitude of $\rm
Kp=20.99\pm0.08$.  Using the B04 model Einstein radius $R_{\rm
  Ein}=0.24$\arcsec, we further measure an Einstein radius aperture
magnitude $\rm Kp_{,Ein}=22.15\pm0.10$.  We are unable to obtain
robust constraints on either the effective radius or the S\'{e}rsic
index, although we note that in all iterations of our modeling, the
S\'{e}rsic index lies between $n=1$ and $n=2.5$, which suggests either
a lenticular or late-type galaxy morphology.

\begin{figure}
\plotone{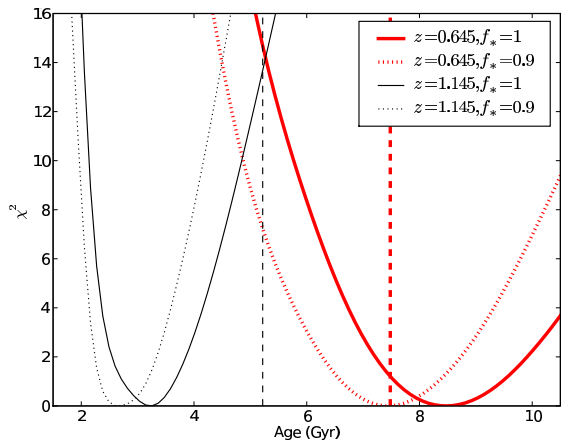}
\caption{Inference on the stellar age of the lensing galaxy by
  comparing the observed magnitude ($\rm Kp=22.15$) with M05 models of
  varying ages. The solid (dotted) curves represent a stellar fraction
  $f_*=1$ ($f_*=0.9$), while
  %while the dotted curves are for $f_*=0.9$ and 
  the vertical dashed lines denote the age of the universe at the
  assumed lens redshift. The two sets of curves represent lens
  redshifts of  $z_l=1.145$ (black) and $z_l=0.645$ (red);
  the lens redshift $z_l=0.218$ leads to stellar population ages older
  than the age of the universe at that redshift for $f_*>0.3$.}
\label{fig:age}
\end{figure}

We use \citet{marast05} instantaneous burst stellar population models
with solar metallicity to infer the age of the lens galaxy stellar
population.  We use models with predominantly stellar lensing mass
(i.e., a stellar fraction $f_*\sim1$) within the Einstein radius,
which are consistent with the results for low-mass S0 galaxies from
\citet{tor09} and with previous results from lenses \citep{aug09}
after extrapolating to the mass of B0128+437. We then infer the
observed Kp-band magnitude for each of the candidate lens redshifts
from the M05 models for a range of stellar population ages and compare
this in a $\chi^2$ sense with the observed magnitude within the
Einstein radius (Figure \ref{fig:age}).  For redshift $z_l=0.218$, no
acceptable solution -- i.e. one in which the stellar age was less than
the age of the universe -- is found unless we assume an unusually low
($f_*=0.3$) stellar fraction.  This analysis does not strongly favor
either of the other two proposed lens redshifts, $z=1.145$ and
$z=0.645$, both of which give stellar ages that are less than the age
of the universe for reasonable values of $f_*$.  However, after
combining this with the previous photometric and spectroscopic data,
we continue to slightly favor the higher redshift $z=1.145$.  The
inferred age for $z=1.145$ is $\sim$3.2 Gyr (Table \ref{tbl:lensb})
and the rest-frame K-band luminosity inferred from the M05 model is
$\log_{10}(L_{\rm K,Ein})=10.61~L_\odot$ (the age and luminosity for
$f_*=0.9$ are 2.7 Gyr and $\log_{10}(L_{\rm K,Ein})=10.67~L_\odot$).
This results in a mass-to-light ratio $M/L_{\rm K,Ein}=0.68$ which is,
by construction, consistent with a stellar population that formed
$\sim$3 Gyr ago.

%--------------------------------------------------------------------
%--------------------------------------------------------------------

\subsection{Flux Ratios}

%--------------------------------------------------------------------
\begin{deluxetable}{cccccc}
\tablecaption{Multi-band Flux-ratio Data \label{tbl:flux}}
\tablehead{
\colhead{Flux Ratio} &\colhead{Kp} & \colhead{X} &\colhead{C} &\colhead{S} &\colhead{L}
}
\startdata
$f_{B}/f_{A}$ &0.17 $\pm$ 0.03  &\nodata  &0.56  &0.49  &\nodata  \\
$f_{C}/f_{A}$ &0.42 $\pm$ 0.01  &\nodata  &0.49  &0.34  &\nodata  \\
$f_{D}/f_{A}$ &0.30 $\pm$ 0.03  &\nodata  &0.47  &0.47  &\nodata  \\
$f_{C}/f_{D}$ &1.43 $\pm$ 0.16  &0.67     &1.04  &0.72  &0.75     \\
\enddata
\tablenotetext{~}{Flux ratios between image pairs.  We present Kp flux
  ratios, along with previously measured flux ratios in the X (3.5
  cm), C (6 cm), S (13 cm), and L (21 cm) radio bands.  In the X and L
  bands, only images C and D are sufficiently deblended to measure
  accurate flux ratios.  Due to the extreme blending of flux between
  images, flux ratios measured in the F160W NICMOS image are
  unreliable, and thus not included.  All radio data are taken from
  B04, with errors $\lesssim$ 5\%.}
\end{deluxetable}
%--------------------------------------------------------------------

Our models give accurate fluxes for each image separately, allowing us
to calculate flux ratios.  We present these flux ratios, along with
the radio flux ratios of B04, in Table \ref{tbl:flux}.  While there is
significant variation between the flux ratios in different bands, one
feature that persists in both the radio and near-IR regimes is the
unusual value of $f_{B}/f_{A}$. The close proximity of A and B
suggests that they should have nearly the same flux and a negligible
time delay \citep[e.g.,][]{kee05}; the discrepancy between what is
seen in the data and what is expected from smooth mass models
therefore cannot be due to intrinsic variability in the source. We
therefore discuss several other phenomena that may contribute to the
flux-ratio anomaly.

\subsubsection{Substructure}
Substructure, in the form of satellite galaxies and dark matter clumps
associated with the lensing galaxy, could explain the anomalous
$f_{B}/f_{A}$, especially since some lenses with flux-ratio anomalies
are known to have luminous satellites whose presence can account for
the anomalies (e.g., MG0414+0534: \citealt{ros00}; MG2016+112:
\citealt{moo09}; B2045+265: \citealt{mck07}).  This scenario is
appealing because VLBI imaging of B0128+437 shows astrometric
anomalies of the quasar image positions.  The presence of substructure
perturbs the gravitational potential and consequently alters the
position \textit{and} the brightness of the lensed images;
substructure could therefore be responsible for both anomalies.

However, since lensing is an achromatic effect, the measured flux
ratios in the radio regime should be the same as those measured in the
near-IR.  Instead, the near-IR $f_{B}/f_{A}$ is significantly smaller
than its radio wavelength counterparts (Table \ref{tbl:flux}).
Therefore, it seems unlikely that substructure alone is responsible
for the anomaly.

Furthermore, after subtracting both the source and lens flux, we find
no obvious luminous substructure in the vicinity of B0128+437; the
flux that remains in the right panel of Figure \ref{fig:imgs} is most
likely an artifact of model subtraction.  However, the fact that we do
not directly detect luminous substructure may not be surprising.  B04
estimate that a perturbing mass as low as $10^6~M_\odot$ could account
for the milliarcsecond-scale astrometric discrepancies observed in
B0128+437.  If we assume reasonable M/L ratios for such an object, it
falls well below our detection threshold.

We determine a limiting magnitude for our imaging by populating the
image with randomly placed simulated point sources covering a range of
magnitudes from $\rm Kp=23.5$ to $\rm Kp=28.5$. We then analyze these
images with SExtractor \citep{ber96} and find that we can repeatedly
recover sources as faint as $\rm Kp=25.5$, which we take to be our
detection limit. This corresponds to a rest-frame V band luminosity
$\rm M_V=-17$, comparable to the Magellanic Clouds. We note, however,
that the detection limit is significantly brighter very close to the
lensed images due to the increased shot noise from the quasars.

\subsubsection{Propagation Effects}
Propagation effects can also account for the $f_{B}/f_{A}$ anomaly,
with scatter broadening by the lensing galaxy interstellar medium
affecting the radio fluxes and dust extinction affecting the near-IR
fluxes, giving rise to the chromatic nature of the flux-ratio anomaly.
If dust were solely responsible for the near-IR anomaly, image B would
need to be dimmed by $\sim$2.0 magnitudes of extinction (assuming as a
lower limit that image A is not being affected).  Using the
\citet{cal00} reddening law for rest-frame J-band, this corresponds to
an E(B-V) value of 1.01, equivalent to $\sim$4.1 magnitudes of
rest-frame V-band extinction.  While this is large, it is not
unphysical: studies of the local group have revealed pockets of dust
in nearby giant molecular clouds (GMCs) that cause up to 10 V-band
magnitudes of extinction \citep{cam99}.

In the radio, scatter broadening by a GMC may explain why image B
shows a decrease in surface brightness and why its subcomponents seen
in 5 GHz VLBI imaging are much broader than those of other images.
This scenario has been postulated for the gravitational lens
B0218+357, where a GMC is thought to be responsible for the
broadening/dimming of image A \citep{gru95,zei10}.

However, while this explains the flux-ratio anomaly, known GMCs are
not massive enough to significantly alter the lensing potential and
cause the astrometric anomalies shown by images B and C.  Therefore,
it is unlikely that propagation effects alone can explain all of the
observed properties.

\subsubsection{Microlensing}
A third phenomenon that could explain the $f_{B}/f_{A}$ flux-ratio
anomaly is extragalactic microlensing, where small gravitational
distortions due to stars in the lensing galaxy affect the fluxes of
images \citep[e.g.,][]{wam06}.  While the size of the radio-emitting
regions of active galaxies are often too large to be significantly
affected by the lensing cross section of individual stars,
microlensing of the accretion disk (the source of the near-IR emission
in B0128+437) can substantially change the observed flux of a given
image. Microlensing could, therefore, explain the discrepant flux
ratios between the radio and near-IR data.

However, while it is possible to observe some microlensing in radio
bands, the flux discrepancies measured for other lens systems are much
less dramatic than the radio $f_{B}/f_{A}$ values of B0128+437.
Studies involving B1608+656: \citep{fas02} and B1600+434:
\citep{koo00} show only small ($\sim$3\%) microlensing radio
flux-ratio anomalies.  Furthermore, based on the movement of stars
relative to the lensed images, the microlensing signal should be
time-variable.  However, radio monitoring of B0128+437 by
\citet{koo03} has shown a nearly time-independent flux ratio.
Finally, microlensing cannot explain the observed astrometric
anomalies.  Thus, it would be difficult for microlensing alone to
fully account for the observed properties of B0128+437.

\subsubsection{Multiple Processes} 
After comparing our data to the existing radio data, we find that no
single phenomenon seems to completely explain the unusual appearance
of B0128+437.  However, since each process is able to explain some
aspect of the anomaly, the full effect is likely due to a combination
of these mechanisms.  Considering the probable late-type morphology of
the lensing galaxy, this multi-process scenario could, for example, be
the result of a spiral arm.  The overdensity of an arm positioned near
image B could be sufficient to cause the astrometric anomaly
\citep{mao98}, and a GMC that is associated with the enhanced star
formation in the spiral arm could produce the wavelength-dependent
flux-ratio anomaly.

However, additional data and modeling are needed before a
definitive explanation for the observed properties of this system can
be made.  Obtaining high-resolution, multi-band imaging in the optical
regime will allow for better constraints on extinction.  Rest-frame U-
and B-band data may also provide visual confirmation of spiral arms,
though given the faintness of the source, deep imaging using either AO
or the JWST will be required to resolve these features.
Similarly, multi-frequency radio observations can be used to probe
scintillation effects such as scatter-broadening.  Higher-resolution
spectroscopy will also be useful.  
For example, if the single emission line associated with the lens
is [\ion{O}{2}], giving $z_l=1.145$, then moderate-resolution
spectroscopy should split the doublet and allow an unambiguous
line identification to be made.
Finally, grid-based modeling of the resolved
Einstein ring flux may provide more robust limits on substructure
\citep{veg09}, allowing us to better quantify the effects of mass
perturbations in this system.

\section{Conclusions}
We use the Keck NIRC2 plus LGS AO to observe the gravitational lens
system B0128+437, and we are able to confirm the presence of the lensing
galaxy and clearly separate the lensing galaxy flux from the source
galaxy flux for the first time.  We take advantage of the high-quality
AO imaging to present new science results for the B0128+437 lens system.
These results can be summarized as follows:

\begin{itemize}

\item For the most probable redshift $z_l=1.145$, we measure a total
  K-band luminosity
  $\log_{10}(L_K)=(11.07\pm0.05)~h_{70}^{-2}~L_{\sun}$, an aperture
  luminosity within the Einstein ring radius $\log_{10}(L_{K,\rm
  Ein})=(10.61\pm0.05)~h_{70}^{-2}~L_{\sun}$, and a K-band M/L ratio
  $\rm{M/L_{K,Ein}}=(0.68\pm0.05)~h_{70}~M_{\sun}/L_{\sun}$ for the
  lensing galaxy.

\item We measure Kp flux ratios of $f_{B}/f_{A}=0.17\pm0.03$,
  $f_{C}/f_{A}=0.42\pm0.01$, and $f_{D}/f_{A}=0.30\pm0.03$.  The
  $f_{B}/f_{A}$ value in the near-IR is significantly smaller than
  both its expected value $f_{B}/f_{A}\sim1$ and its measured radio
  values.  This could be due to the presence of massive substructure,
  propagation effects, or extragalactic microlensing.  We find that
  none of these scenarios alone seems able to fully explain all
  aspects of the anomaly, implying that multiple phenomena are
  responsible.

\item We find that there is no evidence for luminous substructure
  above a point-source limiting magnitude $\rm{Kp}=25.5$,
  corresponding to $\rm{M_V}=-17$ in the B0128+437 field.

\end{itemize}

Looking at the wealth of new information obtained from B0128+437, it
is clear that AO imaging is an important technique for studying
gravitational lenses in new ways.  With future surveys such as the
Joint Dark Enery Mission (JDEM) and Large Synoptic Survey Telescope
(LSST) expected to find $\sim$10000 lenses \citep{mar05}, of which an
estimated 10\% will be of the small-separation, low-mass lensing
galaxy variety \citep{tur84,orb09}, AO imaging of lenses will also
play an important role in analyzing a large subset of future data.

\begin{acknowledgements}
We thank Aaron Dutton for helpful discussions regarding AO observing
and NIRC2 data reduction and Leon Koopmans, John McKean,
Phil Marshall, and Simona Vegetti for useful discussions.  DJL, CDF,
and MWA acknowledge support from NSF-AST-0909119.
The data presented herein were obtained at the W. M. Keck Observatory,
which is operated as a scientific partnership among the California
Institute of Technology, the University of California, and the
National Aeronautics and Space Administration. The Observatory was
made possible by the generous financial support of the W. M. Keck
Foundation.
The authors wish to recognize and acknowledge the very significant
cultural role and reverence that the summit of Mauna Kea has always
had within the indigenous Hawaiian community.  We are most fortunate
to have the opportunity to conduct observations from this mountain.
\end{acknowledgements}

\end{document}